\documentstyle[namedreferences,nato,amsmath,amssymb,10pt]{crckapb}
\begin{opening}
\title{screening and absorption of gravitation in
pre-relativistic and relativistic theories}
\author{H.-H. von Borzeszkowski}
\author{T. Chrobok}
\institute{Institut f\"{u}r Theoretische Physik, Technische Universit\"{a}t
Berlin, Hardenbergstr. 36, D-10623 Berlin, Germany}
\author{H.-J. Treder}
\institute{Rosa-Luxemburg-Str. 17a, D-14482
Potsdam, Germany}

\end{opening}

\runningtitle{}

\begin{document}
\begin{abstract}
After commenting on the early search for a mechanism explaining the
Newtonian action-at-a-distance gravitational law we review
non-Newtonian effects occurring in certain ansatzes for shielding,
screening and absorption effects in pre-relativistic theories of
gravity. Mainly under the aspect of absorption and suppression (or
amplification), we then consider some implications of these ansatzes
for relativistic theories of gravity and discuss successes and
problems in establishing a general framework for a comparison of
alternative relativistic theories of gravity. We examine relativistic
representatives of theories with absorption and suppression (or
amplification) effects, such as fourth-order theories, tetrad
theories and the Einstein-Cartan-Kibble-Sciama theory.
\end{abstract}
\section{Introduction}
All deviations from the gravitational theories of Newton and Einstein
touch fundamental problems of present-day physics and should be
examined experimentally. In particular, such examination provides
further tests of Einstein's general theory of relativity (GRT), which
contains Newton's theory as an approximate case. Therefore, it makes
sense to systematically analyze all effects that differ from the
well-known Newtonian and post-Newtonian ones occurring in GRT (let us
call them non-Einsteinian effects). If these effects can be excluded
experimentally, then this would provide further support for GRT;
otherwise one would have to change basic postulates of present-day
physics. Of all possible non-Einsteinian effects, we shall focus in
this paper on the effects of shielding, absorption and suppression of
gravitation.

The experience made with GRT shows that it is not probable to find
any non-Einsteinian solar-system effects. Nevertheless, it is
useful to search for them on this scale, too, for it can establish
further null experiments in support of GRT. However, a strong
modification of GRT and corresponding effects can be expected on
the microscopic and, possibly, cosmological scale. As to the
microscopic scale, this conjecture is insofar suggesting itself as
there is the problem of quantum gravitation unsolved by GRT. The
cosmological part of this conjecture concerns the standpoint with
respect to the Mach principle. Of course, there is no forcing
physical argument for a realization of this principle, and GRT did
not leave unanswered the question as to it. But one can be
unsatisfied with this answer and look for a generalization of GRT
satisfying this principle more rigorously than GRT. If this is the
case, one has a further argument for a modification of GRT
predicting also non-Einsteinian effects. (In \cite{140}, it was
shown that the tetrad theories discussed in Sec. \ref{TETRAD}
could open up new perspectives for solving both problems
simultaneously.)

We begin in Sec. \ref{EINLEITUNG} with remarks on the early search
for a mechanism explaining the Newtonian action-at-a-distance
gravitational law and then discuss non-Newtonian effects occurring in
certain ansatzes for shielding, screening and absorption effects in
pre-relativistic theories of gravity. In Sec. \ref{HAUPT}, mainly
under the aspect of absorption and suppression (or amplification), we
consider successes and problems in establishing a general framework
for a comparison of alternative relativistic theories of gravity. We
discuss then the following relativistic representatives of theories
with such effects, such as: in Sec. \ref{4ORDNUNG} fourth-order
derivative theories of the Weyl-Lanczos type formulated in Riemann
space-time, in Sec. \ref{TETRAD} tetrad theories formulated in
teleparallelized Riemann space-time, and in Sec. \ref{ECKS} an
effective GRT resulting from the Einstein-Cartan-Kibble-Sciama theory
based on Riemann-Cartan geometry. Finally, in Sec. \ref{ZUSAMMEN} we
summarize our results and compare them with GRT. The resulting
effects are interpreted with respect to their meaning for testing the
Einstein Equivalence Principle and the Strong Equivalence Principle.

\section{Shielding, absorption and suppression in pre-relativistic theories of gravity}
\label{EINLEITUNG}

To analyze non-Einsteinian effects of gravitation it is useful to
remember pre-relativistic ansatzes of the nineteenth century,
sometimes even going back to pre-gravitational conceptions. One
motivation for these ansatzes was to find a mechanical model that
could explain Newton's gravitational inverse-square-law by
something (possibly atomic) that might exist between the
attracting bodies. In our context, such attempts are interesting
to discuss, because they mostly imply deviations from Newton's
law. Another reason for considering such rivals of Newton's law
was that there were several anomalous geodesic, geophysical, and
astronomical effects which could not be explained by Newtonian
gravitational theory. Furthermore, after the foundation of GRT,
some authors of the early twentieth century believed that there
remained anomalies which also could not be explained by GRT. As a
result, pre-relativistic assumptions continued to be considered
and relativistic theories competing with GRT were established.

One influential early author of this story was G.-L. Le Sage
\cite{25,26}. In the eighteenth century, he proposed a mechanical
theory of gravity that was to come under close examination in the
nineteenth century (for details on Le Sage's theory, see also P.
Prevost \cite{38} and S. Aronson \cite{2}). According to this theory,
space is filled with small atomic moving particles which due to their
masses and velocities exert a force on all bodies on which they
impinge. A single isolated body is struck on all sides equally by
these atoms and does not feel any net force. But two bodies placed
next to each other lie in the respective shadows they cast upon each
other. Each body screens off some of the atoms and thus feels a net
force impelling it toward the other body.

Under the influence of the kinetic theory of gases founded in the
1870's Le Sage's theory was revived by Lord Kelvin
\cite{46}\footnote{Later he reconsidered it from the view of
radioactivity \cite{47}.}, S. T. Preston \cite{36,37}, C. Isenkrahe
\cite{21} and P. Drude \cite{13}, bringing Le Sage's hypothesis up to
the standard of a closed theory. However, this approach to
gravitation was rejected by C. Maxwell \cite{32} with arguments
grounded in thermodynamics and the kinetic theory of gases. On the
basis of these arguments, it was discussed critically by Poincar\'{e} and
others (for the English and French part of this early history, cf.
Aronson \cite{2}).

The search for a mechanistic explanation of gravitation and the idea
of a shielding of certain fluxes that intermediate gravitational
interaction were closely related to the question of the accuracy of
Newton's gravitational law. In fact, this law containing only the
masses of the attracting bodies and their mutual distances can only
be exactly valid when neither the intervening space nor the matter
itself absorb the gravitational force or potential. Therefore, it is
not surprising that the possibility of an absorption of gravitation
had already been considered by Newton in his debate with E. Halley
and N. Fatio de Duillier\footnote{Le Sage himself stated that his
speculations go also back to the work of Fatio.}, as is documented in
some of the Queries to Newton's "Opticks." The first research program
looking for an experimental answer to this question was formulated by
M.W. Lomonosov \cite{27} in a letter to L. Euler in 1748. The program
was however only realized 150 years later by R. von E\"{o}tv\"{o}s and Q.
Majorana, without explicit reference to Lomonosov. At about the same
time, Euler discussed with Clairaut, a prominent specialist in
celestial mechanics, the possibility of detecting deviations from
Newton's gravitational law by analyzing the lunar motion. Clairaut
believed for a time that he had found a fluctuation of the lunar
motion testifying to an absorption of gravitation by matter, in this
case, by the earth.

For a long time, the lunar motion has been the strongest criterion
for the validity of the Newtonian and, later, the Einsteinian
theory of gravity (today one would study the motions of artificial
satellites). This is due to the fact that the gravitational
influence of the sun on the moon exceeds the influence of the
earth by the factor 9/4. This solar action varies in dependence on
the distance of the system 'earth-moon' from the sun. Regarding
this effect and, additionally, the action of the other planets on
the lunar motion, a reaming fluctuation of the motion of the moon
could possibly be due to an absorption of solar gravity when the
earth stands between the sun and the moon. This early idea of
Euler was later revived by von Seeliger, and just as Clairaut had
analyzed the lunar motion in order to corroborate it, later
Bottlinger \cite{8,9} did the same in order to find support for
the hypothesis of his teacher von Seeliger \cite{43}.

The first ansatz for an exact mathematical description of
absorption in the sense of Euler and Lomonosov was made by Laplace
\cite{24} in the last volume of his {\it M\'{e}canique
C\'{e}leste}. He assumed that the absorption $d\vec{F}$ of the
flow $\vec{F}$ of the gravitational force is proportional to the
flow $\vec{F}$ itself, the density $\rho$, and the thickness $dr$
of the material penetrated by the gravitational flow, $d\vec{F}=k
\rho\vec{F}dr$. Accordingly a mass element $dm_1$ exerts on
another element $dm_2$ the force
\begin{eqnarray} \label{1}
|d\vec{F}|=\frac{G dm_1 dm_2}{r^2}e^{(-k\rho r)},
\end{eqnarray}
where $k$ is a universal constant of the dimension
$(mass)^{-1}(length)^2$ and $G$ is Newtons gravitational constant.

In the early twentieth century, when Newton's gravitational theory
was replaced by GRT, the two aforementioned attempts by Bottlinger
and Majorana were made to furnish observational and experimental
proof of absorption effects in the sense of Euler and von Seeliger.
Such effects do not exist in GRT and so evidence for them would have
been a blow against the theory.

Using H. von Seeliger's hypothesis of 1909 (von Seeliger \cite{43}),
F.E. Bottlinger \cite{8,9} tried to explain short-period fluctuations
of the motion of the moon (later it became clear that this
explanation was not correct), while Majorana attempted to detect such
absorption effects by laboratory experiments from 1918 till 1930.
Being aware of previous experiments performed to detect an absorption
of gravitation by matter, Majorana turned to this problem in 1918. He
speculated that gravitation was due to a flow of gravitational energy
from all bodies to the surrounding space which is attenuated on
passing through matter. The attenuation would depend exponentially on
the thickness of the matter and its density. Based on a theoretical
estimation of the order of magnitude of this effect he carried out
experiments the results of which seemed to confirm the occurrence of
gravitational absorption. According to present knowledge, they must
have been erroneous (for details of the history of these experiments,
see, e.g., Crowley et al. \cite{11}, Gillies \cite{18}, Martins
\cite{1}).

Another conception competing with absorption or shielding of
gravitation by matter also goes back to papers by von Seeliger
\cite{42}. In these papers, now for cosmological reasons, he
considered a modification of Newton's law by an exponential factor.
Similar ideas were proposed by C. Neumann \cite{34}. At first sight,
it would appear to be the same modification as in the former cases.
This is true, however, only insofar as the form of the gravitational
potential or force is concerned. For the differential equations to
which the respective potentials are solutions, there is a great
difference. In the first case, instead of the potential equation of
Newtonian theory,
\begin{eqnarray} \label{0NEU}
\bigtriangleup\phi=4\pi G\rho,
\end{eqnarray}
one has equations with a so-called potential-like coupling,
\begin{eqnarray} \label{2}
\bigtriangleup\phi=4\pi G\rho \phi,
\end{eqnarray}
while in the second case one arrives at an equation with an
additional vacuum term,
\begin{eqnarray} \label{3}
\bigtriangleup\phi-k^2\phi=4\pi G\rho.
\end{eqnarray}
The latter equation requires the introduction of a new fundamental
constant $k$ corresponding to Einstein's cosmological constant
$\Lambda$. As will be seen in Sec. \ref{TETRAD}, in relativistic
theory a combination of vacuum and matter effects can occur, too.

In \cite{50} it is shown that, regarding cosmological consequences,
both approaches are equivalent. Indeed, if one replaces in (\ref{2})
$\phi$ by $\phi+c^2$ and considers a static cosmos with an average
matter density $\bar{\rho}=const$ then the corresponding average
potential $\bar{\phi}$ satisfies the equation
\begin{eqnarray} \label{4NEU}
\bigtriangleup\bar{\phi}-k^2\bar{\phi}=4\pi G\bar{\rho}
\end{eqnarray}
where
\begin{eqnarray} \label{5NEU}
k^2=4\pi G\bar{\rho}c^{-2}
\end{eqnarray}
which is identical to the averaged equation (\ref{3}).

Equation (\ref{2}) shows that the potential-like coupling of matter
modelling the conception of an absorption of the gravitational flow
by the material penetrated can also be interpreted as a dependence of
the gravitational number on the gravitational potential and thus on
space and time. Therefore, to some extent it models Dirac's
hypothesis within the framework of a pre-relativistic theory, and a
relativistic theory realizing Dirac's idea could have equation
(\ref{2}) as a non-relativistic approximation. Another possible
interpretation of (\ref{2}) is that of a suppression of gravitation
(self-absorption) by the dependence of the active gravitational mass
on the gravitational potential. Indeed, the product of the matter
density and the gravitational potential can be interpreted as the
active gravitational mass.

Not so much for historical reasons but for later discussion, it is
interesting to confront the potential equations (\ref{0NEU}) and
(\ref{3}) with the bi-potential equations \cite{74},
\begin{eqnarray} \label{8NEU}
\bigtriangleup\bigtriangleup\phi=4\pi G\rho \\ \label{9NEU}
\bigtriangleup(\bigtriangleup-k^2)\phi=4\pi G\rho
\end{eqnarray}  or, for a point-like distribution of matter described by
the Dirac delta-function $\delta$,
\begin{eqnarray} \label{10NEU}
\bigtriangleup\bigtriangleup\phi=4\pi a \delta(\vec{r}) \\
\label{11NEU} \bigtriangleup(\bigtriangleup-k^2)\phi=4\pi a
k^2\delta(\vec{r}).
\end{eqnarray}
Eqs. (\ref{10NEU}) and (\ref{11NEU}) show that the elementary
solutions are given by the Green functions,
\begin{eqnarray} \label{12NEU}
\phi=\frac{ar}{2}\\ \label{13NEU} \phi=\frac{a}{r}-\frac{a}{r}e^{-kr}
\end{eqnarray}
indicating that there is a long-range (Newtonian) and a short-range
(self-absorption) part of gravitational interaction.

In order to select these physically meaningful solutions from a
greater manifold of solutions for the fourth-order equations one had
to impose yet an additional condition on the structure of sources:
The distribution of matter must be expressed by a monopole density.
This becomes obvious when one considers the general
spherical-symmetric vacuum solutions of (\ref{0NEU}, \ref{11NEU})
which are given by
\begin{eqnarray} \label{14NEU}
\phi=\frac{ar}{2}+b+\frac{c}{r}+\frac{d}{r^2} \\ \label{15NEU}
\phi=\frac{A}{r}+B+C\frac{e^{-kr}}{r}+D\frac{e^{kr}}{r}
\end{eqnarray}
respectively, satisfying the equations
\begin{eqnarray} \label{16NEU}
\bigtriangleup\bigtriangleup\phi=-4\pi a \delta(\vec{r})-4\pi
c\bigtriangleup\delta(\vec{r}) \\ \label{17NEU}
\bigtriangleup(\bigtriangleup-k^2)\phi=-4\pi(A+C)\bigtriangleup
\delta(\vec{r})-4\pi k^2C\delta(\vec{r}).
\end{eqnarray}

Originally, such equations where discussed in the framework of the
Bopp-Podolsky electrodynamics. This electrodynamics states that,
for $A=-C$, there exist two kinds of photons, namely massless and
massive photons. They satisfy equations which, in the static
spherical-symmetric case, reduce to Eq. (\ref{9NEU}). It was shown
in \cite{IVANENKO} that its solution (\ref{13NEU}) is everywhere
regular and $\phi(0)=ak$ ($a=e$ is the charge of the electron).
The same is true for gravitons: A theory with massless and massive
gravitons requires fourth order equations (see Sec.
\ref{4ORDNUNG}).\footnote{In \cite{TREDER} it was shown that it is
erroneous to assume that massive gravitons occur in Einsteins GRT
with a cosmological term.}

The atomic hypotheses assuming shielding effects lead in the static
case to a modification of Newton's gravitational law that is
approximately given by the potential introduced by von Seeliger and
Majorana. Instead of the $r^2$-dependence of the force between the
attracting bodies given by the Newtonian fundamental law,
\begin{eqnarray} \label{4}
\vec{F}_{12}=-G\int\frac{\rho(\vec{r}_1)\rho(\vec{r}_2)}{r_{12}^3}\vec{r}_{12}d^3x_1d^3x_2
\end{eqnarray}
where $G$ is the Newtonian gravitational constant, one finds,
\begin{eqnarray} \label{5}
\vec{F}_{12}=-G\int\frac{\rho(\vec{r}_1)\rho(\vec{r}_2)}{r_{12}^3}\vec{r}_{12}e^{(-k\int\rho
dr_{12})} d^3x_1d^3x_2.
\end{eqnarray}
Here the exponent $-k\int\rho dr_{12}$ means an absorption of the
flow of force $\vec{F}$  by the atomic masses between the two
gravitating point masses. Since, for observational reasons, one has
to assume that the absorption exponent is much smaller than $1$, as a
first approximation, (\ref{5}) may be replaced by Laplace's
expression,
\begin{eqnarray} \label{6}
\vec{F}_{12}^*=-G\frac{M_1M_2}{r_{12}^3}\vec{r}_{12}\exp/(-k\rho
dr)\approx -G\frac{M_1M_2}{r_{12}^3}\vec{r}_{12}(1-k\rho \triangle
r).
\end{eqnarray}

Equation (\ref{6}) contains a new fundamental constant, namely
Majorana's "absorption coefficient of the gravitational flow"
\begin{eqnarray} \label{7}
k\geq 0, [k]=cm^2 g^{-1}.
\end{eqnarray}
This value can be tested by the E\"{o}tv\"{o}s experiment, where one can
probe whether the ratio of the gravitational and the inertial mass of
a body depends on its physical properties. In the case of absorption
of gravitation the value of this ratio would depend on the density of
the test body. Gravimetric measurements of the gravitational constant
carried out by E\"{o}tv\"{o}s by means of a torsion pendulum and
gravitational compensators showed that $k$ has to be smaller than
\begin{eqnarray} \label{8a}
k<4 {\times}10^{-13}cm^2 g^{-1}.
\end{eqnarray}
By comparison Majorana \cite{28,29,30} obtained in his first
experiments the value
\begin{eqnarray} \label{8b}
k\approx 6.7{\times} 10^{-12}cm^2 g^{-1},
\end{eqnarray}
which was compatible with his theoretical analysis. (Later, after
some corrections, he arrived at about half this value \cite{31}).

A more precise estimation of $k$ can be derived from
celestial-mechanical observations. As mentioned above, Bottlinger
hypothesized that certain (saros-periodic) fluctuations of the motion
of the moon are due to an absorption of solar gravity by Earth when
it stands between Sun and Moon\footnote{A. Einstein commented
Bottlinger's theory in \cite{15,16,17} (see also \cite{51}).}. If we
assume this hypothesis, then, following Crowley et al. \cite{11}, the
amplitude $\lambda$ of these fluctuations is related to the
absorption coefficient $k$ via
\begin{eqnarray} \label{9}
\lambda\approx 2k a\rho,
\end{eqnarray}
where $\rho$ denotes the mean density and $a$ the radius of the
earth. If one assumes that
\begin{eqnarray} \label{10}
k\approx6.3{\times} 10^{-15}cm^2 g^{-1},
\end{eqnarray}
then the value of $\lambda$ is in accordance with the so-called great
empirical term of the moon theory. This, however, also shows that, if
the fluctuations of the motion of moon here under consideration had
indeed been explained by von Seeliger's absorption hypothesis, then
greater values than the one given by (\ref{10}) are not admissible as
they would not be compatible with the motion theory. That there is
this celestial-mechanical estimation of an upper limit for $k$ had
already been mentioned by Russell \cite{40} in his critique of
Majorana's estimation (\ref{8b}).

A better estimate of $k$ has been reached by measurements of the
tidal forces. According to Newton's expression, the tidal force
acting upon earth by a mass $M$ at a distance $R$ is
\begin{eqnarray} \label{11}
Z=-2\frac{GM}{R^2}\frac{a}{R}
\end{eqnarray}
where $a$ denotes the radius of the earth. However von Seeliger's
and Majorana's ansatz (\ref{5}) provided
\begin{eqnarray} \label{12}
Z^*\approx -2\frac{GM}{R^2}\frac{a}{R}-\frac{\lambda
GM}{2R^2}\approx-2\frac{GM a}{R^3}(1+\frac{\lambda R}{4 a})
\end{eqnarray}
with the absorption coefficient of the earth body
\begin{eqnarray} \label{12a}
\lambda\approx 2 k a \rho\approx 6.6{\times}10^9 cm^{-2} g \times k.
\end{eqnarray}

Considering now the ratio of the tidal forces due to the sun and
moon, $Z_s$ and $Z_m$, one finds in the Newtonian case
\begin{eqnarray} \label{13}
\frac{Z_s}{Z_m}\approx\frac{M_s}{M_m}\left(\frac{R_m}{R_s}\right)^3=\frac{5}{11}
\end{eqnarray}
and in the von Seeliger-Majorana case
\begin{eqnarray} \label{14}
\frac{Z_s^*}{Z_m^*}\approx\frac{M_s}{M_m}\left(\frac{R_m}{R_s}\right)^3
(1+\lambda\frac{R_s}{a})=\frac{5}{11}(1+4 k \times 10^{13}cm^{-2}g).
\end{eqnarray}
Measurements carried out with a horizontal pendulum by Hecker
\cite{20} gave the result
\begin{eqnarray} \label{15a}
\frac{Z_s^*}{Z_m^*}\leq1, \quad \mbox{that is,}\quad k<2{\times}
10^{-14}g^{-1}cm^2.
\end{eqnarray}
This result is still compatible with Bottlinger's absorption
coefficient, but not with Majorana's value (\ref{8b}), which
provided a sun flood much greater than the moon flood (Russell
\cite{40}).

Later, Hecker's estimation was confirmed by Michelson and Gale
\cite{33}, who by using a "level" obtained
\begin{eqnarray} \label{15b}
\frac{Z_s^*}{Z_m^*}=0.69\pm 0.004, \quad \mbox{i. e.,} \quad k<1.3{\times}
10^{-14}g^{-1}cm^2.
\end{eqnarray}
(The real precision of these measurements, however, was not quite
clear.)

Bottlinger \cite{8,9} had also proposed to search for jolting
anomalies in gravimeter measurements occurring during solar eclipses
due to a screening of the gravitational flow of the sun by the moon.
In an analysis performed by Slichter et al. \cite{44}, however, this
effect could not be found and those authors concluded that $k$ has
the upper limit $3{\times}10^{-15}cm^2g^{-1}$. However, as argued earlier
\cite{5}, measurements of this effect provide by necessity null
results due to the equivalence of inertial and passive gravitational
masses verified by E\"{o}tv\"{o}s.

The latest observational limits on the size of the absorption
coefficient is $k < 10^{-21} cm^2g^{-1}$. It was established by a
reanalysis of lunar laser ranging data (Eckardt \cite{14}, cf.
also Gillies \cite{19}). This would rule out the existence of this
phenomenon, at least in the way that it was originally envisioned.
(For an estimation of that part which, from the viewpoint of
measurement, is possibly due to shielding effects, cf. \cite{55}.)
The actual terrestrial experimental limit provides $k < 2\times
10^{-17} cm^2g^{-1}$ \cite{UNNI01}.

About the same estimate follows from astrophysics \cite{45,51}.
Indeed, astrophysical arguments suggest that the value for $k$ has to
be much smaller than $10^{-21} cm^2 g^{-1}$. This can be seen by
considering objects of large mass and density like neutron stars. In
their case the total absorption can no longer be described by the
Seeliger-Majorana expression. However, one can utilize a method
developed by Dubois-Reymond (see Drude \cite{13}) providing the upper
limit $k =3/R\rho$, where $R$ is the radius and $\rho$ the density of
the star. Assuming an object with the radius $10^6 cm$ and a mass
equal to $10^{34} g$ one is led to $k = 10^{-22} cm^2 g^{-1}$.

As in the aforementioned experiments, these values for $k$ exclude
an absorption of gravitation in accordance with the
Seeliger-Majorana model. But it does not rule out absorption
effects as described by relativistic theories of gravity like the
tetrad theory, where the matter source is coupled potential-like
to gravitation \cite{5}. The same is true for other theories of
gravity competing with GRT that were systematically investigated
as to their experimental consequences in Will \cite{54}. For
instance, in the tetrad theory, the relativistic field theory of
gravity is constructed such that, in the static non-relativistic
limit, one has (for details see Sec \ref{TETRAD})
\begin{eqnarray} \label{16}
\bigtriangleup\phi=4\pi G\rho\left(1-\frac{\alpha
|\phi|}{c^2r_{12}}\right), \quad \alpha=const.
\end{eqnarray}
From this equation it follows that there is a suppression of
gravitation by another mass or by its own mass. In the case of two
point masses, the mutual gravitational interaction is given by
\begin{eqnarray} \nonumber
m_1\ddot{\vec{r}}_1=-\frac{G m_1
m_2}{r_{12}^3}\vec{r}_{12}\left(1-\frac{2Gm_1}{c^2r_{12}}\right),
 \\ \label{17}  m_2\ddot{\vec{r}}_2=-\frac{G m_1
m_2}{r_{12}^3}\vec{r}_{12}\left(1-\frac{2Gm_2}{c^2r_{12}}\right).
\end{eqnarray}
Thus the effective active gravitational mass $m$ is diminished by the
suppression factor $(1-2Gm/c^2r_{12})$. Such effects can also be
found in gravitational theories with a variable 'gravitational
constant' (Dirac \cite{12}, Jordan \cite{22,23}, Brans and Dicke
\cite{10}). Furthermore, in the case of an extended body one finds a
self-absorption effect. The effective active gravitational mass
$\bar{M}$ of a body with Newtonian mass $M$ and radius $r$ is
diminished by the body's self-field,
\begin{eqnarray} \label{18}
\bar{M}=GM\left(1-\frac{4 \pi G}{3c^2}\rho r^2\right)
\end{eqnarray}
where exact calculations show that the upper limit of this mass is
approximately given by the quantity $c^2/\sqrt{G\rho}$.

The modifications of the Newtonian law mentioned above result from
modifications of the Laplace equation. In their relativistic
generalization, these potential equations lead to theories of
gravity competing with GRT. On one hand GRT provides, in the
non-relativistic static approximation, the Laplace equation and
thus the Newtonian potential and, in higher-order approximations,
relativistic corrections. On the other hand, the competing
relativistic theories lead, in the first-order approximation, to
the above mentioned modifications of the Laplace equation and
thus, besides the higher-order relativistic corrections, to
additional non-Newtonian variations. All these relativistic
theories of gravity (including GRT) represent attempts to extend
Faraday's principle of the local nature of all interactions to
gravitation. Indeed, in GRT the geometrical interpretation of the
equivalence principle realizes this principle insofar as it
locally reduces gravitation to inertia and identifies it with the
local world metric; this metric replaces the non-relativistic
potential and Einstein's field equations replace the Poisson's
potential equation. Other local theories introduce additional
space-time functions which together with the metric describe the
gravitational field. In some of the relativistic rivals of GRT,
these functions are of a non-geometric nature. An example is the
Jordan or Brans-Dicke scalar field, which, in accordance with
Dirac's hypothesis, can be interpreted as a variable gravitational
'constant' $G$. (For a review of theories involving absorption and
suppression of gravitation, see \cite{5}.)  In other theories of
gravity, these additional functions are essentials of the
geometric framework, as for instance tetrad theories working in
teleparallel Riemannian space and the metric-affine theories
working in Riemann-Cartan space-times that are characterized by
non-vanishing curvature and torsion (see Secs \ref{TETRAD},
\ref{ECKS}).

\section{On absorption, self-absorption and suppression of gravitation in relativistic
theories} \label{HAUPT}

The original idea of shielding gravitation assumed that the insertion
of some kind of matter between the source of gravitation and a test
body could reduce the gravitational interaction. This would be a
genuine non-Einsteinian effect. However, in the present paper, our
main concern will be relativistic theories. Those theories satisfy
the Einstein Equivalence Principle (EEP) \cite{5,54} stating the
following:
\begin{itemize}
\item[\textbf{i})] The trajectory of an uncharged test body depends only on the
initial conditions, but not on its internal structure and
composition, and
\item[\textbf{ii})] the outcome of any non-gravitational experiment
is independent of the velocity of the freely falling apparatus and of
the space-time position at which it is performed (local Lorentz
invariance and local position invariance).
\end{itemize}
The point is that, as long as one confines oneself to theories
presupposing the EEP, shielding effects will not occur. For, this
principle implies that there is only one sign for the gravitational
charge such that one finds quite another situation as in
electrodynamics where shielding effects appear ("Faraday screen").
But other empirical possibilities for (non-GRT) relativistic effects
like absorption and self-absorption are not excluded by the EEP.

In the case of absorption the intervening matter would not only
effect on the test body by its own gravitational field but also by
influencing the gravitational field of the source. If the latter
field is weakened one can speak of absorption. (For a discussion
of this effect, see also \cite{56}). In contrast to absorption,
self-absorption describes the backreaction of the gravitational
field on its own source so that the effective active gravitational
mass of a body (or a system of bodies) is smaller than its
uneffected active gravitational mass. Of course, both absorption
and self-absorption effects can occur simultaneously, and
generally both violate the Strong Equivalence Principle (SEP)
\cite{5,54} stating:
\begin{itemize}
\item[\textbf{i})]
The trajectories of uncharged test bodies as well as of
self-gravitating bodies depend only on the initial conditions, but
not on their internal structure and composition, and
\item[\textbf{ii})]
the outcome of any local test experiment is independent of the
velocity of the freely falling apparatus and of the space-time
position at which it is performed.
\end{itemize}
GRT fulfils SEP. The same is true for fourth-order metric theories of
gravity, although there one has a suppression (or amplification) of
the matter source by its own gravitational field (see Sec.
\ref{4ORDNUNG}).

To empirically test gravitational theories, in particular GRT, it is
helpful to compare them to other gravitational theories and their
empirical predictions. For this purpose, it was most useful to have a
general parameterized framework encompassing a wide class of
gravitational theories where the parameters, whose different values
characterize different theories, can be tested by experiments and
observations. The most successful scheme of this type is the
Parameterized Post-Newtonian (PPN) formalism \cite{54} confronting
the class of metric theories of gravity with solar system tests.

This formalism rendered it possible to calculate effects based on
a possible difference between the inertial, passive and active
gravitational masses. It particularly was shown that some theories
of gravity predict a violation of the first requirement of the SEP
that is due to a violation of the equivalence between inertial and
passive gravitational masses. This effect was predicted by Dicke
\cite{57} and calculated by Nordtvedt \cite{58}. It does not
violate the EEP. Other effects result from a possible violation of
the equivalence of passive and active gravitational masses. For
instance, Cavendish experiments could show that the locally
measured gravitational constant changes in dependence on the
position of the measurement apparatus. This can also be
interpreted as a position-dependent active gravitational mass,
i.e., as absorption.

The PPN formalism is based on the four assumptions:
\begin{itemize}
\item[\textbf{i})]
slow motion of the considered matter,
\item[\textbf{ii})]
weak gravitational fields,
\item[\textbf{iii})]
perfect fluid as matter source, and
\item[\textbf{iv})]
gravitation is described by so-called metric theories (for the
definition of metric theories, see below).
\end{itemize}
As noticed in \cite{54}, due to the first three assumptions, this
approach is not adequate to compare the class of gravitational
theories described in (iv) with respect to compact objects and
cosmology. And, of course, in virtue of assumption (iv), the
efficiency of this approach is limited  by the fact that it is
dealing with metric theories satisfying EEP.

During the last decades great efforts were made to generalize this
framework to nonmetric gravitational theories and to reach also a
theory-independent parameterization of effects which have no
counterpart in the classical domain. (For a survey, see e. g.
\cite{59,60}.) The generalization to a nonmetric framework enables
one to test the validity of the EEP, too, and the analysis of
"genuine" quantum effects leads to new test possibilities in
gravitational physics, where these effects are especially appropriate
to test the coupling of quantum matter to gravitation.

First of all, this generalized framework is appropriate to analyze
terrestrial, solar system, and certain astrophysical experiments
and observations. Of course, it would be desirable to have also a
general framework for a systematic study of gravitational systems
like compact objects or cosmological models in alternative
theories of gravity. But one does not have it so that one is
forced to confine oneself to case studies what, to discuss
relativistic absorption effects, in the next section will be done.

Our considerations focus on theories that use the genuinely
geometrical structures metric, tetrads, Weitzenb\"{o}ck torsion, and
Cartan torsion for describing the gravitational field. Accordingly,
the following case studies start with theories formulated in a
Riemannian space, go then over to theories in a teleparallelized
space, which is followed by a simple example of a theory established
in a Riemann-Cartan space\footnote{Of course, there are other
alternatives to GRT which are also interesting to be considered with
respect to non-Einsteinian effects, such as scalar-tensor theories of
Brans-Dicke type and effective scalar-tensor theories resulting from
higher-dimensional theories of gravity by projection,
compactification or other procedures (for such theories, see also the
corresponding contributions in this volume). At first the latter idea
was considered by Einstein and Pauli \cite{EIN27,113}. They showed
that, if a Riemannian $V_5$ with signature $(-3)$ possesses a Killing
vector which is orthogonal to the hypersurface $x^4=const$, $V_5$
reduces to a $V_4$ with a scalar field because then one has $(A,
B=0,\ldots,4)$: $g_{AB,4}=0, g_{i4}=0,
g_{44}=(g^{44})^{-1}=g_{44}(x^l)$.
\\
 It should be
mentioned that the fourth-order theories discussed in Sec. 3.1 and
the scalar-tensor theories are interrelated (cf., the references to
equivalence theorems given in Sec.3.1). Furthermore, there is also an
interrelation between fourth-order theories, scalar-tensor theories,
and theories based on metric-affine geometry which generalizes
Riemann-Cartan geometry by admitting a non-vanishing non-metricity.
First this became obvious in Weyl's theory \cite{WEY18} extending the
notion of general relativity by the requirement of conformal
invariance and in Bach's interpretation of Weyl's "unified" theory as
a scalar-tensor theory of gravity \cite{91} (see also, \cite{73}).}.

But before turning to this consideration, a remark concerning the
relation between EEP and metric theories of gravity should be made.
This remark is motivated by the fact that it is often argued that the
EEP necessarily leads to the class of metric theories satisfying the
following postulates:
\begin{itemize}
\item[\textbf{i})]
Spacetime is endowed with a metric $\mathbf{g}$,
\item[\textbf{ii})]
the world lines of test bodies are geodesics of that metric, and
\item[\textbf{iii})]
in local freely falling frames, called local Lorentz frames, the
non-gravitational physics are those of special relativity.
\end{itemize}    (Therefore, presupposing the validity of EEP,
in \cite{54} only such theories are compared.)

The arguments in favor of this thesis are mainly based on the
observation (see, e.g., \cite{54}, p.22 f.) that one of the
aspects of EEP, namely the local Lorentz invariance of
non-gravitational physics in a freely falling local frame, demands
that there exist one or more second-rank tensor fields which
reduce in that frame to fields that are proportional to the
Minkowski metric $\mathbf{\eta}$. Another aspect of EEP, local
position invariance then shows that the scalar factor of the
Minkowski metric is a universal constant from which the existence
of a unique, symmetric, second-rank tensor field $\mathbf{g}$
follows. However, the point we want to make here is that this does
not automatically mean that the gravitational equations have to be
differential equations for the metric (and, possibly, additional
fields, as for instance a scalar field which do not couple to
gravitation directly, as, e.g. realized in the Brans-Dicke theory
\cite{10}). This conclusion can only be drawn if one assumes that
the metric is a primary quantity. Of course, in physically
interpretable theory, one needs a metric and from EEP it follows
the above-said. But one can also consider gravitational theories,
wherein other quantities are primary. Then one has field equations
for these potentials, while the metric describing, in accordance
with EEP, the influence of gravitation on non-gravitational
matter, is a secondary quantity somehow derived from the primary
potentials. For instance, such theories are tetrad (or
teleparallelized) theories \cite{62,63,64} (see also
\cite{5,140,7}) or affine theories of gravity
\cite{67,68,69,70,71} (see also \cite{72,73}).

\subsection{Fourth-order theories} \label{4ORDNUNG}
Fourth-order derivative equations of gravitation have received great
attention since several authors have proved that the fourth-order
terms $R^2$, $R_{ik}R^{ik}$, and $R_{iklm}R^{iklm}$ can be introduced
as counter-terms of the Einstein-Hilbert Lagrangian $R$ to make GRT,
quantized in the framework of covariant perturbation theory, one-loop
renormalizable. In this context it was clarified that
higher-derivative terms naturally appear in the quantum effective
action describing the vacuum polarization of the gravitational field
by gravitons and other particles \cite{87,88,89,90}. Such equations
revive early suggestions by Bach \cite{91}, Weyl \cite{92,93},
Einstein \cite{94}, and Eddington \cite{95} (see also Pauli \cite{96}
and Lanczos \cite{97,971}).

While early papers treated such equations formulated in
non-Riemannian spacetime in order to unify gravitational and
electromagnetic fields\footnote{Higher-derivative equations were also
studied within the framework of quantum field theory by Pais and
Uhlenbeck \cite{98}.}, later they were also considered as classical
theories. From the viewpoint of classical gravitational equations
with phenomenological matter modifying the Einstein gravitation at
small distances, the discussion of such field equations was opened by
Buchdahl \cite{99} and Pechlaner and Sexl \cite{100}. Later this
discussion was anew stimulated by the argument \cite{101,102} that
such equations should be considered in analogy to the fourth-order
electromagnetic theory of Bopp and Podolsky \cite{103,104} in order
to solve the singularity and collapse problems of GRT. Especially, it
was shown \cite{101,102} that, for physical reasons, the
Einstein-Hilbert part of the Lagrangian must be necessarily included
(for this point, see also \cite{105}) and that one has to impose
supplementary conditions on the matter source term. These
considerations were continued in \cite{74,75} (see also
\cite{106,107,108}). The consequences for singularities of massive
bodies and cosmological collapse were also analyzed in
\cite{109,110}. (In the following, we shall call fourth-order
equations supplemented by the second-order Einstein term {\it mixed
fourth-order equations}.)

The meaning of such higher order terms in early cosmology was
considered in \cite{120,121}. Especially the influence of pure
$R^2$-terms in cosmological scenarios can be interpreted as a pure
gravitational source for the inflationary process \cite{117,118,122}.
Moreover it was shown in \cite{119} that the transformation
$\tilde{g}_{ab}=(1+2\epsilon R)g_{ab}$ leads back to GRT described by
$\tilde{g}_{ab}$ with an additional scalar field $R$ acting as a
source. (For more general equivalence theorems, see \cite{SOKO94} and
the corresponding references cited therein.)

In the case of stellar objects in the linearized, static, weak field
limit for perfect fluid the corresponding Lane-Emden equation
provides a relative change of the radius of the object compared to
Newtons theory depending on the value of the fundamental coupling
constants \cite{123}. This may decrease or increase the size of the
object.

Mixed fourth-order equations may only be considered as gravitational
field equations, if they furnish approximately at least the
Newton-Einstein vacuum for large distances. To this end, it is not
generally sufficient to demand that the static spherical-symmetric
solutions contain the Schwarzschild solution the Newtonian potential
in the linear approximation. One has to demand that the corresponding
exact solution can be fitted to an interior equation for physically
significant equations of state. In the case of the linearized field
equations, this reduces to the requirement that there exist solutions
with suitable boundary conditions which, for point-like particles,
satisfy a generalized potential equation with possessing a delta-like
source.

The most general Lagrangian containing the Hilbert-Einstein invariant
as well as the quadratic scalars reads as follows\footnote{In this
section, we mainly follow \cite{74,75}.} (the square of the curvature
tensor need not be regarded since it can be expressed by the two
other terms, up to a divergence):
\begin{eqnarray} \label{54}
L=\sqrt{-g}R+\sqrt{-g}(\alpha R_{ab}R^{ab}+\beta R^2)l^2+2 \kappa L_M
\end{eqnarray}
where $\alpha$ and $\beta$ are numerical constants, $l$ is a constant
having the dimension of length. Variation of the action integral
$I=\int L d^4x$ results in the field equations of the fourth order,
viz.,
\begin{eqnarray} \label{55}
l^2H_{ab}+E_{ab}:=\\ \nonumber =l^2\bigl(\alpha \Box
R_{ab}+(\frac{\alpha}{2}+2\beta)g_{ab} \Box
R-(\alpha+2\beta)R_{;ab}+2\beta RR_{ab}-\\ \nonumber
-\frac{\beta}{2}g_{ab}R^2+2\alpha
R_{acdb}R^{cd}-\frac{\alpha}{2}g_{ab}R_{cd}R^{cd}\bigr)+\left(R_{ab}-\frac{1}{2}Rg_{ab}\right)=-\kappa
T_{ab}.
\end{eqnarray}
(Here $\Box$ denotes the covariant wave operator.) These equations
consist of two parts, namely, the fourth-order terms $\propto l^2$
stemming from the above quadratic scalars and the usual Einstein
tensor, where
\begin{eqnarray} \label{56}
T_{ab}:=\frac{1}{\sqrt{-g}}\frac{\delta L_M}{\delta g^{ab}} \quad
(\mbox{with} \quad T^a{}_{b;a}=0).
\end{eqnarray}

For $\alpha=-2\beta=1$ (assumed by Eddington in the case of pure
fourth-order equations), in the linear approximation the vacuum
equations corresponding to Eqs. (\ref{55}) reduce to
\begin{eqnarray}
l^2\Box_{\eta} E^1_{ab}+E^1_{ab}=0
\end{eqnarray}
and, for  $\alpha=-3\beta=1$ (assumed by Bach and Weyl in the case
of pure fourth-order equations), to
\begin{eqnarray}
l^2\Box_{\eta} R^1_{ab}+R^1_{ab}=0
\end{eqnarray}
(the latter follows because in this case the trace of the vacuum
equations provides $R=0$). The discussion of these equations
performed in \cite{74} shows that both cases are characterized in
that they possess, besides massless gravitons, only {\it one} kind of
gravitons with non-vanishing restmass. (These results were
corroborated by subsequent considerations \cite{106,108}.)

In the linear approximation for static fields with the Hilbert
coordinate condition
\begin{eqnarray}
g^{ik}{}_{,k}=\frac{1}{2}g^k{}_{k,i}
\end{eqnarray}
in the above-considered two cases, the pure and mixed fourth order
field equations rewritten in a compact form as
\begin{eqnarray}
l^2H_{ik}=-\kappa T_{ik},\\ l^2H_{ik}+E_{ik}=-\kappa T_{ik},
\end{eqnarray}
produce the simple potential equations (\ref{10NEU}) and
(\ref{11NEU}) for all components of the metric tensor, if
$\alpha=-3\beta$ or $\alpha=-2\beta$, where $k\propto l^{-1}$ is a
reciprocal length, and $a$ is the mass of a point-like particle which
is given by the Dirac delta-function $\delta$. As shown in Sec.
\ref{EINLEITUNG}, it is this condition imposed on the source that
leads to the Green functions (\ref{12NEU}) and (\ref{13NEU}). From
the expressions (\ref{12NEU}) and (\ref{14NEU}) it is evident that,
in the case of pure field equations of fourth order, this condition
will reduce the manifold of solutions to a functions that do not
satisfy the correct boundary conditions at large distances, i.e.,
those field equations do not mirror the long-range Newtonian
interaction. They are consequently at most field equations describing
free fields, in other words, equations of a unified field theory in
the sense of Weyl's \cite{92} and Eddington's \cite{95} ansatz.

The requirement of general coordinate covariance furnishes in the
fourth-order case a variety of field equations depending upon the
parameters $\alpha$ and $\beta$, whereas the Einstein equations are
determined by this symmetry group up to the cosmological term. If one
postulates that the field equations are invariant with respect to
conform transformations, one obtains just the pure fourth-order
equations of Bach and Weyl, where $\alpha=-3\beta$. However, if one
wants to couple gravitation to matter one has to go a step further
since, due to the conform invariance, the trace of $H_{ik}$ vanishes.
Accordingly, the trace of the matter must also vanish. That means
that in this case there are supportive reasons for the replacement of
pure by mixed field equations. Then the conform invariance is broken
by the term $E_{ik}$ being the source of massless gravitons
\cite{74}.

The meaning of the short-range part of the gravitational potential
of this theory becomes more evident when one has a glance at the
stabilizing effect it has in the classical field-theoretical
particle model (for more details, see \cite{75}). Within GRT, such
models first were tried to construct by Einstein \cite{111},
Einstein and Rosen \cite{112}, and Einstein and Pauli \cite{113}.
Later this work was continued in the frame of the so-called geon
program by Wheeler and co-workers. Already in the first paper by
Einstein dealing with this program it was clear that, in the
Einstein-Maxwell theory, it was difficult to reach a stable model.
In particular, it was shown that a stable model requires to assume
a cut-off length and the equality of mass and charge. The fact
that the situation in regard to the particle problem becomes more
tractable under these two conditions is an indication that GRT
should be modified, if one wants to realize this program. As far
as very dense stars are concerned, such a modification should
already become significant if the distances are of the order of
magnitude of the gravitational radii.

Indeed, such a modification is given by the mixed field equations.
In the linear approximation given by Eq. (\ref{11NEU}), one is led
to the Podolsky type solution (Greek indices run from $1\dots3$),
\begin{eqnarray}
g_{00}=1+2\phi, \quad g_{0\alpha}=0, \quad
g_{\alpha\beta}=-\delta_{\alpha\beta}(1-2\phi)
\end{eqnarray}
where
\begin{eqnarray}
\phi=\frac{GM}{r c^2}(1-e^{-kr}).
\end{eqnarray}
By calculating the affine energy-momentum tensor $t^i{}_k$ of the
gravitational field with this solution and integrating $t^i{}_k$ over
a spatial volume, one obtains the result that the Laue criterion
\cite{114} for a stable particle with finite mass yields:
\begin{eqnarray}
\int t^0{}_0d^3x=\frac{GM^2}{2}c^2k, \quad \int
t^0{}_{\alpha}d^3x=\int t^{\alpha}{}_0d^3x=0.
\end{eqnarray}
This character of the Podolsky potential can also be seen if one
introduces it in the Einstein model \cite{115} of a stellar object
(see \cite{116}).

The left-hand pure vacuum part of the field equations (\ref{55}) with
arbitrary $\alpha$ and $\beta$ are constructed according to the same
scheme as Einstein's second-order equations. They are determined by
the requirement that they are not to contain derivatives higher than
fourth order and fulfil the differential identity $H^k{}_{i;k}=0$. By
virtue of this identity and the contracted Bianchi identity for
$E_{ik}$, $E^k{}_{i;k}=0$, we get, as in the case of Einstein's
equations, the dynamic equations
\begin{eqnarray}
T^k{}_{i;k}=0,
\end{eqnarray}
for the pure and mixed equations of fourth order. Consequently, the
EEP is also satisfied for those equations.

As to the SEP, the fourth-order metric theory satisfies all
conditions of this principle formulated above. But this does not
exclude a certain back-reaction of the gravitational field on its
matter source leading to the fact that the effective active
gravitational mass differs from the inertial mass. This can made
plausible by the following argument \cite{107}: Regarding that for
vacuum solutions of Eqs. (\ref{55}) satisfying the asymptotic
condition
\begin{eqnarray}
g_{00}=1-\frac{2a}{r} \quad \mbox{for} \quad r\rightarrow\infty
\end{eqnarray}
the active gravitational mass is given by (see \cite{113})
\begin{eqnarray}
a=\int \int\int (E^0{}_0-E^{\alpha}{}_{\alpha})\sqrt{-g}d^3x
\end{eqnarray}
we obtain
\begin{eqnarray}
a=\kappa \int \int\int
(T^0{}_0-T^{\alpha}{}_{\alpha})\sqrt{-g}d^3x-l^2 \int \int\int
(H^0{}_0-H^{\alpha}{}_{\alpha})\sqrt{-g}d^3x.
\end{eqnarray}
This relation replaces the GRT
equation
\begin{eqnarray}
a=\kappa \int \int\int
(T^0{}_0-T^{\alpha}{}_{\alpha})\sqrt{-g}d^3x.
\end{eqnarray}
It represents a modification of the Einstein-Newtonian equivalence
of active gravitational mass $a$ and inertial mass $m$ stemming
from the hidden-matter term $\propto l^2$. It behaves like hidden
(or "dark") matter since, due to the relation $H^{ik}{}_{;k}=0$,
$H^{ik}$ decouples from the visible-matter term $T^{ik}$. (In the
case $\alpha=-3\beta$, this term may be interpreted as a second
kind of gravitons). In dependence on the sign of this term, it
suppresses or amplifies the active gravitational mass.

\subsection{Tetrad theories} \label{TETRAD}
That class of gravitational theories which leads to a
potential-like coupling is given by tetrad theories formulated in
Riemannian space-time with teleparallelism. To unify
electromagnetism and gravitation it was introduced by Einstein and
elaborated in a series of papers, partly in cooperation with Mayer
(for the first papers of this series, see \cite{76,77,78,79,80}).
From another standpoint, later this idea was revived by M\o ller
\cite{62,63} and Pellegrini \& Pleba\'nski \cite{64}, where the
latter constructed a general Lagrangian based on Weitzenb\"{o}ck's
invariants \cite{84}. These authors regard all $16$ components of
the tetrad field as gravitational potentials which are to be
determined by corresponding field equations. The presence of a
non-trivial tetrad field can be used to construct, beside the
Levi-Civita connection defining a Riemannian structure with
non-vanishing curvature and vanishing torsion, a teleparallel
connection with vanishing curvature and non-vanishing torsion.
This enables one to build the Weitzenb\"{o}ck invariants usable as
Lagrangians. (Among them, there is also the tetrad equivalent of
Einstein-Hilbert Lagrangian, as was shown by M\o ller.)

A tetrad theory of gravity has the advantage to provide a
satisfactory energy-momentum complex \cite{62,64}. Later M\o ller
\cite{63} considered a version of this theory that can free
macroscopic matter configurations of singularities.

From the gauge point of view, in \cite{MULL82} tetrad theory was
regarded as translational limit of the Poincar\'e gauge field theory,
and in \cite{MULL83} such a theory was presented as a constrained
Poincar\'e gauge field theory\footnote{For Poincar\'e gauge field
theory, see Sec. \ref{ECKS}.}. Another approach \cite{82} considers
the translational part of the Poincar\'e group as gauge group, where
in contrast to Poincar\'e gauge field theory this theory is assumed
to be valid on microscopic scales, too. A choice of the Lagrangian
leading to a more predictable behavior of torsion than in the
above-mentioned versions of the tetrad theory is discussed in
\cite{83} - from the point of Mach's principle, tetrad theory was
considered in \cite{140}.

The general Lagrange density which is invariant under global Lorentz
transformations and provides differential equations of second order
is given by
\begin{eqnarray} \label{35}
L^*=\sqrt{-g}(R+aF_{Bik}F^{Bik}+b\Phi_A\Phi^A)+2\kappa L_M
\end{eqnarray}
where $a$, $b$ are constants, $R=g^{ik}R_{ik}$ is the Ricci scalar,
$L_M$ denotes the matter Lagrange density and
\begin{eqnarray} \label{36}
F_{Aik}:=h_{Ak,i}-h_{Ai,k}=h_A{}^l(\gamma_{lki}-\gamma_{lik}),
\quad \Phi_A:=h_A{}^i\gamma^m{}_{im}.
\end{eqnarray}
Here $h_A{}^l$ denote the tetrad field and
$\gamma_{lki}=h^A{}_lh_{Ak;i}$ the Ricci rotation coefficients
($A,B,\dots$ are tetrad (anholonomic) indices, $i,k, \dots$ are
space-time (holonomic) indices).

To consider absorption and self-absorption mechanisms we confine
ourselves to the case $b=0$, such that the Lagrange density takes the
form
\begin{eqnarray} \label{37}
L=\sqrt{-g}(R+aF_{Bik}F^{Bik})+2\kappa L_M.
\end{eqnarray}
(Einstein \cite{76} and Levi-Civita \cite{84} discussed the
corresponding vacuum solution as a candidate for a unified
gravito-electrodynamic theory.) Introducing the tensors
\begin{eqnarray} \label{38}
T_{ik}:=\frac{1}{\sqrt{-g}}\frac{\delta L_M}{\delta h^i{}_A}h^A{}_k\\
\label{39} G_{ik}:=R_{ik}-\frac{1}{2}g_{ik}R+\kappa
T_{ik}+2a(\frac{1}{4}g_{ik}F_{Bmn}F^{Bmn}-F_{Bim}F^{B}{}_k{}^m)
\end{eqnarray}
and the $4$-vector densities
\begin{eqnarray} \label{40}
S_{A}{}^i=\sqrt{-g}h_A{}^kG^i{}_k
\end{eqnarray}
and varying the Lagrangian (\ref{37}) with respect to the tetrad
field, it follow the gravitational field equations in the "Maxwell"
form
\begin{eqnarray} \label{41}
F_{A}{}^{ik}{}_{;k}=\frac{1}{2a}S_A{}^i.
\end{eqnarray}
These equations can also be rewritten in an "Einstein" form. This
form stems from a Lagrangian which differs from (\ref{37}) by a
divergence term \cite{TRE78}. Since, in the following paragraphs of
this section, we assume a symmetric energy-momentum tensor this form
be given for this special case:
\begin{eqnarray} \label{42}
R_{ik}-\frac{1}{2}g_{ik}R&=&-\kappa T_{ik}-\Theta_{(ik)}\\ \label{43}
\Theta_{ik}-\Theta_{ki}&=&0
\end{eqnarray}
where
\begin{eqnarray} \label{44}
\Theta_{ik}=a(\frac{1}{2}g_{ik}F_{Bmn}F^{Bmn}-F_{Bim}F^{B}{}_k{}^m+
2h^A{}_iF_{Ak}{}^l{}_{;l}).
\end{eqnarray}

The latter form of the field equations allows for an interesting
interpretation, because the purely geometric term (\ref{44}) can be
regarded as matter source term\footnote{If one considers the vacuum
version of (\ref{42}) in the context of unified geometric field
theory this interpretation is even forcing (see Schr\"{o}dinger
\cite{71}).}. Furthermore, for a symmetric tensor $T_{ik}$, due to
the dynamical equations and Bianchi's identities, one has \cite{140}
\begin{eqnarray} \label{45}
\Theta_{i}{}^k{}_{;k}=0
\end{eqnarray}
so that $\Theta_{ik}$ does not couple to visible matter described by
$T_{ik}$. Again it behaves like hidden (or "dark") matter.

The Maxwell form (\ref{41}) shows clearly that this theory can be
considered as a general-relativistic generalization of Eq. (\ref{2}).
Interestingly, in this form the gravitational (suppressing or
amplifying) effect appears as a combination of an absorption effect
given by the potential-like coupling of usual matter and a
hidden-matter effect. Of course, both versions of the theory must
lead to the same empirical results.

To discuss absorption effects following from the potential-like
coupling in more detail, let us consider the absorption of the active
gravitational mass of Earth by the gravitational field of Sun, i.e.,
calculate the change of the spherical-symmetric part of the Earth
field in dependence on the position in the Sun field. To do this, we
shall follow a method used in \cite{86} for equations similar to
(\ref{41}) (see also \cite{5}). This effect gives an impression of
the order of magnitude of the effects here under consideration.

To this end, we consider the field of Earth, $\underset{1}{h}^A{}_i$,
as a perturbation of the field of Sun, $\underset{0}{h}^A{}_i$:
\begin{eqnarray} \label{46}
h^A{}_i=\underset{0}{h}^A{}_i+\underset{1}{h}^A{}_i \quad \mbox{
with} \quad |\underset{1}{h}^A{}_i|\ll |\underset{0}{h}^A{}_i|.
\end{eqnarray}
Rewriting Eq. (\ref{41}) in the form
\begin{eqnarray} \label{47}
\Box h^A{}_i-h^{Ak}{}_{,ik}=\frac{h_{,k}}{h}F^A{}_i{}^k+
\frac{1}{2a}h^A{}_lE^l{}_{i}+\frac{\kappa}{2a}h^A{}_lT^l{}_{i}+\\
\nonumber +\frac{1}{4}h^A{}_iF_{Bmn}F^{Bmn}-h^{Al}F_{Blm}F^B{}_i{}^m
\end{eqnarray}
(where $h:=\sqrt{-g}=det|h^A{}_i|$ and
$E_{ik}=R_{ik}-\frac{1}{2}g_{ik}R$) and inserting ansatz (\ref{46})
one obtains in the first order approximation
\begin{eqnarray} \label{48}
\Box \underset{0}{h}^A{}_i+\Box \underset{1}{h}^A{}_i-
\underset{0}{h}^{Ak}{}_{,ik}-\underset{1}{h}^{Ak}{}_{,ik}=
\frac{\underset{0}{h}{}_{,k}}{\underset{0}{h}}\underset{0}{F}^A{}_i{}^k+
\frac{\underset{0}{h}{}_{,k}}{\underset{0}{h}}\underset{1}{h}\underset{0}{F}^A{}_i{}^k
+\underset{1}{h}{}_{,k}\underset{0}{F}^A{}_i{}^k+\\ \nonumber +
\frac{\underset{0}{h}{}_{,k}}{\underset{0}{h}}\underset{1}{F}^A{}_i{}^k
+\frac{1}{2a}\underset{0}h^A{}_l\underset{1}E^l{}_i+
\frac{1}{2a}\underset{1}h^A{}_l\underset{0}E^l{}_i+
\frac{1}{2a}\underset{0}h^A{}_l\underset{0}E^l{}_i+
\frac{1}{4}\underset{0}h^A{}_l\underset{0}{F}{}_{Bmn}\underset{0}{F}^{Bmn}
+\\ \nonumber +\frac{1}{4}(h^A{}_iF_{Bmn}F^{Bmn})_1-
\underset{0}h^{Al}\underset{0}{F}{}_{Blm}\underset{0}{F}^B{}_i{}^m
-(h^{Al}F_{Blm}F^B{}_i{}^m)_1+\\ \nonumber
+\frac{\kappa}{2a}\left(\underset{0}{h}^A{}_l\underset{E}T^l{}_i
+\underset{1}{h}^A{}_l\underset{S}T^l{}_i+
\underset{1}{h}^A{}_l\underset{E}T^l{}_i+\underset{0}{h}^A{}_l\underset{S}T^l{}_i\right),
\end{eqnarray}
where $\underset{E}T^l{}_i$ and $\underset{S}T^l{}_i$ denote the
energy-momentum tensor of Earth and Sun, respectively, and
$\sqrt{-g}:=h=\underset{0}{h}+\underset{1}{h}$. Regarding that the
solar field $\underset{0}{h}^A{}_i$ is a solution of that equation
which is given by the zero-order terms in Eq. (\ref{48}) one gets as
first-order equation for $\underset{1}{h}^A{}_i$
\begin{eqnarray} \label{49}
\Box
\underset{1}{h}^A{}_i-\underset{1}{h}^{Ak}{}_{,ik}=\frac{\kappa}{2a}\left(\underset{0}{h}^A{}_l\underset{E}T^l{}_i
+\underset{1}{h}^A{}_l\underset{S}T^l{}_i+
\underset{1}{h}^A{}_l\underset{E}T^l{}_i\right)+\\ \nonumber +
\frac{\underset{0}{h}{}_{,k}}{\underset{0}{h}}\underset{1}{h}\underset{0}{F}^A{}_i{}^k
+\underset{1}{h}{}_{,k}\underset{0}{F}^A{}_i{}^k+
\frac{\underset{0}{h}{}_{,k}}{\underset{0}{h}}\underset{1}{F}^A{}_i{}^k+\frac{1}{2a}\underset{0}h^A{}_l\underset{1}E^l{}_i+\\
\nonumber + \frac{1}{2a}\underset{1}h^A{}_l\underset{0}E^l{}_i
+\frac{1}{4}(h^A{}_iF_{Bmn}F^{Bmn})_1 -(h^{Al}F_{Blm}F^B{}_i{}^m)_1.
\end{eqnarray}
Now one can make the following assumptions
\begin{itemize}
\item[\textbf{i})]{The terms in the second and third lines all contain
cross-terms in $\underset{0}{h}^A{}_l$ and $\underset{1}{h}^A{}_l$.
They describe, in the first-order approximation, the above-mentioned
hidden matter correction of the active gravitational mass. In the
following we assume that the constant $a$ is so small that the usual
matter terms in the first line of (\ref{49}) are dominating.}
\item[\textbf{ii})]{The term $\underset{1}{h}^A{}_l\underset{S}T^l{}_i$ is neglected.
It describes the influence of Earth potential on the source of the
solar field. Near the Earth, it causes a correction having no
spherical-symmetric component with respect to the Earth field.}
\item[\textbf{iii})]{The energy-momentum tensor is assumed to have
 the form $\underset{E}T^{ik}=\rho u^i u^k$ (with $\rho=const$).
A more complicated ansatz regarding the internal structure of the
Earth in more detail leads to higher-order corrections.}
\item[\textbf{iv})]{The term
$\underset{1}{h}^A{}_l\underset{E}T^l{}_i$ is neglected, too. It
describes the influence of the Earth field on its own source leading
to higher-order (self-absorption) effects.}
\end{itemize}
As a consequence of these assumptions, Eqs. (\ref{49}) reduce to the
field equations
\begin{eqnarray} \label{50}
\Box
\underset{1}{h}^A{}_i-\underset{1}{h}^{Ak}{}_{,ik}=\frac{\kappa}{2a}\underset{0}{h}^A{}_l\underset{E}T^l{}_i
\end{eqnarray}

Assuming coordinates, where the spherical-symmetric solution
$\underset{1}{h}^A{}_i$ has the form ($\mu, \nu=1,2,3$),
\begin{eqnarray} \label{51}
\underset{1}{h}^{\hat{0}}{}_0=\alpha, \quad
\underset{1}{h}^{\hat{\mu}}{}_{\nu}=\delta^{\hat{\mu}}_{\nu}\beta,
\quad\underset{1}{h}^{\hat{0}}{}_{\nu}=\underset{1}{h}^{\hat{\nu}}{}_0=0,
\end{eqnarray}
Eqs. (\ref{50}) lead to the equations
\begin{eqnarray} \label{52}
\bigtriangleup\alpha=-\frac{\kappa}{2a}\underset{0}{h}^0{}_lT^l{}_0\\
\nonumber
\bigtriangleup\beta=-\frac{\kappa}{4a}\underset{0}{h}^{\mu}{}_{\nu}T^{\nu}{}_{\mu}.
\end{eqnarray}
Therefore, up to a factor, the calculation of the absorption effect
provides the same result as given in \cite{85}. Thus a gravimeter
would register an annual period in the active gravitational mass of
the Earth depending on the distance Earth-Sun. The mass difference
measured by a gravimeter at aphelion and perihelion results as
\begin{eqnarray} \label{53}
\frac{\bigtriangleup M}{M}=\frac{4 G
M_{\odot}}{c^2R}\epsilon\frac{1}{2a}
\end{eqnarray}
($M_{\odot}=$ is the mass of the Sun, $R=$ the average distance of
the planet from the Sun, and $\epsilon=$ the eccentricity of the
planetary orbit). For Earth, this provides the value
$6.6\times10^{-10}$ and, for Mercury $2.06\times10^{-8}$ if $2a=1$.

It should be mentioned that the above-made estimation is performed
under the assumption that the planet rests in the solar field (what
enables one to assume (\ref{51})). However, one has to regard that
the square of the velocity of the Earth with respect to the Sun is of
the same order of magnitude as the gravitational field of the Sun.
Thus, the velocity effect could compensate the absorption effect. As
was shown in \cite{86}, the velocity effect is of the same order of
magnitude, but generally it differs from the absorption effect by a
factor of the order of magnitude $1$.

\subsection{Einstein-Cartan-Kibble-Sciama theory} \label{ECKS}
The Einstein-Cartan-Kibble-Sciama theory \cite{1191,1201,1211} (see
also, for an early ansatz \cite{1221} and for a later detailed
elaboration \cite{1231}) is the simplest example of a Poincar\'e
gauge field theory of gravity, aiming at a quantum theory of gravity.
Moreover, the transition to gravitational theories that are
formulated in Riemann-Cartan space is mainly motivated by the fact
that then the spin-density of matter (or spin current) can couple to
the post-Riemannian structure.\footnote{For modern motivation and
representation of metric-affine theories of gravity, see
\cite{HEHL}.} Therefore, the full content of this theory becomes only
obvious when one considers the coupling of gravitation to spinorial
matter.

As far as the non-quantized version of the theory is concerned, there
exist two elaborated descriptions of spinning matter in such a
theory, the classical Weyssenhoff model \cite{Weyssenhoff,KOOB1987}
and a classical approximation of the second quantized Dirac equation
\cite{1251}. Moreover there is a number of interesting cosmological
solutions which show that in such a theory it is possible to prevent
the cosmological singularity \cite{KOPZ,KUCH}. But, unfortunately
this is not a general property of Einstein-Cartan-Kibble-Sciama
cosmological solutions \cite{KOOB1987}.

In the following we shall describe the situation of an effective
Einstein-Cartan-Kibble-Sciama theory with Dirac-matter as source
term. This theory
is formulated in the Riemann-Cartan space $U_4$ \cite{1241}, where
the non-metricity $Q_{ijk}$ is vanishing, $Q_{ijk}:=-\nabla_i
g_{jk}=0$, such that the connection is given as:
\begin{eqnarray} \label{72}
\Gamma^k_{ij}=\left\{ {k \atop ij } \right\} -K_{ij}{}^k
\end{eqnarray}
with the Christoffel symbols $\left\{ {k \atop ij } \right\}$, the
torsion $S_{ij}{}^k$, and the contorsion $K_{ij}{}^k$, which in the
holonomic representation read,
\begin{eqnarray} \label{73}
\left\{ {k \atop ij }
\right\}:=\frac{1}{2}g^{kl}(g_{li,j}+g_{jl,i}-g_{ij,l}),
\\ \nonumber
S_{ij}{}^k:=\frac{1}{2}(\Gamma^k_{ij}-\Gamma^k_{ji}), \quad
K_{ij}{}^k:=-S_{ij}{}^k+S_{j}{}^k{}_{i}-S^{k}{}_{ij}=-K_{i}{}^k{}_{j}.
\end{eqnarray}
In the anholonomic representation, where all quantities are referred
to an orthonormal tetrad of vectors
$\overrightarrow{e}_A=e^i_A\partial_i$ (capital and small Latin
indices run again from $0$ to $3$)\footnote{In contrast to Sec.
\ref{TETRAD}, we denote the tetrad with $e$ instead of $h$ since here
they are coordinates in the Riemann-Cartan space but not a fixed
tetrad field in Riemann space with teleparallelism.}, the pure metric
part $ $ of the anholonomic connection $\tilde{\Gamma}$ is given by
the Ricci rotation coefficients $\gamma_{AB}{}^i$.

The field equations are deduced by varying the action integral
corresponding to a Lagrange density that consists of a pure
gravitational and a matter part; in the holonomic version one has
\begin{equation} \label{74a} L=L_G(g,\partial g,\Gamma,
\partial \Gamma)+2\kappa L_M(g,
\partial g, \Gamma, \phi, \partial \phi)
\end{equation}
and, in the anholonomic version,
\begin{equation}\label{74b}
L=L_G(e, \partial e, \tilde{\Gamma}, \partial \tilde{\Gamma})+2
\kappa L_M(e,
\partial e, \tilde{\Gamma}, \phi, \partial \phi).
\end{equation}
In the holonomic representation, one has to vary (\ref{74a}) with
respect to
\begin{center} $g_{ij}$ and $\Gamma^k_{ij}$ or, for
$\Gamma=\Gamma(K,g)$, $g_{ij}$ and $K_{ij}{}^k$
\\ \hspace{2cm}
or, for $K=K(S,g)$, $g_{ij}$ and $S_{ij}{}^k$
\end{center}
and, in the anholonomic representation, (\ref{74b}) with respect to
\begin{center}
$e^i_A$ and $\tilde{\Gamma}_{AB}{}^i$ or, for
$\tilde{\Gamma}=\tilde{\Gamma}(K,\gamma)$, $e^i_A$ and
$K_{AB}{}^i$.
\end{center}
If matter with half-integer spin is to be coupled to gravitation,
then one has to transit to the spinorial version of the latter, where
one has to vary
\begin{center}
$\gamma^i=e^i_A\gamma^A$ and $\omega_{AB}{}^i$ (or $S_{AB}{}^i$)
\end{center}
($\gamma^A$ denote the Dirac matrices defined by
$\{\gamma^A,\gamma^B\}=2\eta^{AB}$ and $\omega_{AB}{}^i$ the spinor
connection and $S_{AB}{}^i$ the spinor torsion).

In the case of the Kibble-Sciama theory the pure gravitational
Lagrangian is assumed to be the Ricci scalar such that $L_G=eR$ (with
$e:=\det(e^A_i)$, $e^A_i$ is dual to $e^i_A$). Then the variation of
(\ref{74b}) by $e^i_A$ and $\omega_{iAB}$ provides the gravitational
equations\footnote{Here we follow the (slightly modified) notations
used in \cite{1251}. They differ from those ones in \cite{1231} by
sign in $R_{ik}$ and $T_{ikl}$, and spin density defined in
\cite{1251} is twice that one in \cite{1231} (Therefore, it appear in
\cite{1241} and our formulas (\ref{75}) and (\ref{76}) the minus
signs and in Eq. (\ref{76}) the factor $1/2$).}:
\begin{eqnarray}\label{75}
R^A{}_i-\frac{1}{2}e^A{}_iR=-\kappa T^A{}_i,\\ \label{76}
T_{ikl}=-\frac{\kappa}{2}s_{ikl}
\end{eqnarray}
where
\begin{eqnarray} \label{77}
R^A{}_i-\frac{1}{2}e^A{}_iR:=e^k{}_BR_{ik}{}^{AB}-\frac{1}{2}e^A{}_ie^l{}_Ce^m{}_DR_{lm}{}^{CD}\\
\label{78} T_{ikl}:=S_{ikl}-2\delta^l_{[i}S_{k]m}{}^m, \\
\label{79} T_i{}^A:=-\frac{1}{e}\frac{\delta L_M}{\delta e^i{}_A},
\\ \label{80} s^{ikl}:=-\frac{1}{e}\frac{\delta L_M}{\delta
\omega_{iAB}}e^k{}_Ae^l{}_B.
\end{eqnarray}


Assuming the case of a Dirac matter field, where $L_M$ is given by
the expression,
\begin{eqnarray} \label{81}
L_D=\hbar
c(\frac{i}{2}\overline{\psi}(\gamma^i\nabla_i-\overleftarrow{\nabla}_i\gamma^i)\psi-m\overline{\psi}\psi),
\end{eqnarray}
the source terms read (see, e.g., \cite{1231} and \cite{1251}),
\begin{eqnarray} \label{82}
T^A{}_i=\frac{\hbar
c}{2}\overline{\psi}(i\gamma^A\nabla_i-\overleftarrow{\nabla}_i i
\gamma^A)\psi, \\ \label{83} s^{ikl}=s^{[ikl]}=\frac{\hbar
c}{2}\overline{\psi}\gamma^{[i}\gamma^k\gamma^{l]}\psi= \frac{\hbar
c}{2}\epsilon^{ikls}\overline{\psi}\gamma_s\gamma_5\psi,
\end{eqnarray}
where $\epsilon^{ikls}$ is the Levi-Civita symbol and
$\gamma_5:=i\gamma^0\gamma^1\gamma^2\gamma^3$.

Due to the total anti-symmetry of the spin density and the field
equations (\ref{78}), the modified torsion $T_{ikl}$ and the torsion
$S_{ikl}$ itself are also completely anti-symmetric. Therefore,
because of (\ref{73}), one has $\Gamma^k_{(ij)}=\left\{ {k \atop ij }
\right\}$. (It should be mentioned that, for the total antisymmetry
of $S_{ikl}$, the Dirac equations following by varying the Lagrangian
(\ref{81}) with respect to $\psi$ and $\overline{\psi}$ couple only
to metric but not to torsion (see \cite{1251}).)

Because of the second field equation (\ref{78}) one can substitute
the spin density for the torsion and this way arrive at effective
Einstein equations. For this purpose, we split the Ricci tensor into
the Riemannian and the non-Riemannian parts ($\overset{0}{}$ denotes
Riemannian quantities and $_;$ the covariant derivative with respect
to $\{ \}$),
\begin{eqnarray} \label{84}
R_{Bi}=e^l_AR^A{}_{Bli}=\\ \nonumber
=e^l_A(\overset{0}{R}{}^A{}_{Bli}-S^A{}_{Bl;i}+S^A{}_{Bi;l}-S^C{}_{Bl}S^A{}_{Ci}+S^C{}_{Bi}S^A{}_{Cl})=\\
\nonumber =\overset{0}{R}_{Bi}+S^l{}_{Bi;l}.
\end{eqnarray}
Regarding that the Ricci scalar is given as
\begin{eqnarray} \label{85}
R=e^l_Ae^{Bi}R^A{}_{Bli}=\overset{0}{R}+S_{CAD}S^{ACD}
\end{eqnarray}
we obtain Einstein tensor
\begin{eqnarray} \label{86}
R_{Bi}-\frac{1}{2}e_{Bi}R=\overset{0}{R}_{Bi}-\frac{1}{2}e_{Bi}\overset{0}{R}+S^l{}_{Bi;l}-\frac{1}{2}e_{Bi}S_{CAD}S^{ACD}
\end{eqnarray}
such that the effective Einstein equations have the form
\begin{eqnarray} \label{87}
\overset{0}{R}_{Bi}-\frac{1}{2}e_{Bi}\overset{0}{R}-\frac{\kappa^2}{2}e_{Bi}s_{klm}s^{klm}=\\
\nonumber =-\kappa\left(\overset{0}{T}{}^{D}_{Bi}-\kappa\frac{i\hbar
c}{8}\overline{\psi}\gamma^k\sigma^{lm}s_{klm}\psi e_{Bi}
-\frac{1}{2}s^l{}_{Bi;l}\right)
\end{eqnarray}
($\overset{0}{T}{}^{D}_{Bi}$ is the energy-momentum tensor of the
Dirac field in the Riemannian space and
$\sigma^{lm}=[\gamma^l,\gamma^m]$.) These equations differ by $3$
terms from the Einstein equations:
\begin{itemize}
\item[\textbf{i})]{On the left-hand side, it appears a
cosmological term, where the effective cosmological constant
$\Lambda$ in the epoch under consideration, is given by the product:
$1/2\times$ square of gravitational constant $\times$ square of the
spin density of matter in this epoch. Due to the structure of
$s_{klm}$ given by Eq. (\ref{83}), this term can be rewritten as
follows\footnote{$\psi^+=\overline{\psi}^*$ (complex-conjugate and
transposed $\psi$)}
\begin{eqnarray} \label{88}
-\frac{\kappa^2}{2}s_{klm}s^{klm}=-\frac{\kappa^2}{8}(\epsilon_{klms}\overline{\psi}\gamma^s\gamma_5\psi)(\epsilon^{klmt}\overline{\psi}\gamma_t\gamma_5\psi)=\\
\nonumber
=\frac{3\kappa^2}{8}\delta^t_s\psi^+\gamma_5\gamma^s\gamma_5\psi\psi^+\gamma_5\gamma_t\gamma_5\psi=\\
\nonumber
=\frac{3\kappa^2}{8}\psi^+\gamma^s\gamma_s\psi|\psi|^2=\frac{3\kappa^2}{2}|\psi|^4
\end{eqnarray}
where $|\psi|^2=\psi^+\psi$. In other words, for a finite fermion
number, the effective cosmological term is proportional to the square
of the fermion density. Thus, a cosmological term can be dynamically
induced or, if such a term is assumed to exist at the very beginning
in the equations, tuned away in an effective
Einstein-Cartan-Kibble-Sciama theory. (For other approaches, there is
a vast number of papers that one has to regard. For the approach of
effective scalar-tensor theories, see, e. g., \cite{1261} and the
literature cited therein.)}
\item[\textbf{ii})]{On the right-hand side, it appears an additional
source term given by the spin density $s_{klm}$ and a divergence
term. Originally, the latter is also a pure geometric term, namely
the divergence of the torsion $-S^{l}{}_{Bi;l}$. It has the same
property as the dark-matter terms in fourth-order metric theory (Sec.
\ref{4ORDNUNG}) and the tetrad theory (Sec. \ref{TETRAD}):
\begin{eqnarray} \label{89}
S^l{}_{Bi;l}{}^i=-g_{si}S_B{}^{ls}{}_{;l}{}^i=-S_B{}^{ls}{}_{;ls}=0.
\end{eqnarray}
But, in contrast to the above-discussed cases, in virtue of the
second field equation (\ref{78}), it can be rewritten in a
visible-matter term.}
\end{itemize}

There is still another way to go over from the
Einstein-Cartan-Kibble-Sciama theory to an effective GRT. Instead
of combining the field equations (\ref{75}) and (\ref{76}), this
way performs the substitution $S_{ikl}\rightarrow\omega_{ikl}$ in
the Lagrangian
\begin{eqnarray} \label{90}
L(e^A_i,\partial e^a_i,S_{iAB},\partial
S_{iAB},\psi,\partial\psi,\overline{\psi},\partial\overline{\psi})=\\
\nonumber =e R(e^A_i,\partial e^A_i,S_{iAB},\partial
S_{iAB})+L_D(e^A_i,\partial
e^A_i,S_{iAB},\psi,\partial\psi,\overline{\psi},\partial\overline{\psi})
\end{eqnarray}
where $L_D$ is given by Eq. (\ref{81}). This provides the effective
Lagrangian
\begin{eqnarray} \label{91}
L^*(e^A_i,\partial
e^A_i,\psi,\partial\psi,\overline{\psi},\partial\overline{\psi})=
eR^*(e^A_i,\partial e^A_i)+\\ \nonumber +2\kappa L^*_D(e^A_i,\partial
e^A_i,S_{iAB},\psi,\partial\psi,\overline{\psi},\partial\overline{\psi})\\
\label{92} R^*(e^A_i,\partial
e^A_i)=\overset{0}{R}-\omega_{ikl}\omega^{ikl}=\\ \nonumber
=\overset{0}{R}-S_{ikl}S^{ikl}=
\overset{0}{R}-\frac{\kappa^2}{4}s_{ikl}s^{ikl}=\overset{0}{R}+\frac{3\kappa^2}{2}|\psi|^4\\
\label{93} L^*_D(e^A_i,\partial
e^A_i,S_{iAB},\psi,\partial\psi,\overline{\psi},\partial\overline{\psi})=
\overset{0}{L}_D+\frac{1}{4}i\hbar
ce(\overline{\psi}\gamma^i\sigma^{AB}\omega_{iAB}\psi)=\\ \nonumber
=\overset{0}{L}_D+\frac{1}{4}i\hbar
ce(\overline{\psi}\gamma^i\sigma^{AB}S_{iAB}\psi)=\\ \nonumber
=\overset{0}{L}_D-\frac{1}{8}i\hbar
ce(\overline{\psi}\gamma^i\sigma^{AB}s_{iAB}\psi).
\end{eqnarray}

The variation of $L^*$ with respect to $e^A_i$ provides the
equations,
\begin{eqnarray} \label{94}
\overset{0}{R}_{Bi}-\frac{1}{2}e_{Bi}\overset{0}{R}-
\frac{3\kappa^2}{2}e_{Bi}|\psi|^4=-2\kappa\frac{1}{e}\frac{\delta
L^*_D}{\delta e^B_k}g_{ki}
\end{eqnarray}
where
\begin{eqnarray} \label{95}
\frac{1}{e}\frac{\delta L^*_D}{\delta
e^A_i}=(\overset{0}{T}{}^D)^i_A-\frac{1}{8}\frac{\kappa i \hbar
c}{e}\frac{\delta e}{\delta
e^A_i}(\overline{\psi}\gamma^c\sigma^{AB}s_{cAB}\psi)=\\ \nonumber
=(\overset{0}{T}{}^D)^i_A-\frac{1}{8}i \kappa \hbar
c(\overline{\psi}\gamma^c\sigma^{AB}s_{cAB}\psi)e^i_A
\end{eqnarray}
such that , finally, one has:
\begin{eqnarray} \label{96}
\overset{0}{R}_{Bi}-\frac{1}{2}e_{Bi}\overset{0}{R}-\kappa^2e_{Bi}\left(
\frac{3}{2}|\psi|^4-\frac{1}{4}i \hbar
c(\overline{\psi}\gamma^c\sigma^{AD}s_{cAD}\psi)\right)=-\kappa\overset{0}{T}{}^D_{Bi}.
\end{eqnarray}
In contrast to Eq. (\ref{87}), in Eq. (\ref{96}) the term
proportional to $s_{klm}$ is absorbed in the effective cosmological
term, while the "wattless" term is missing. But, generally, one meets
again the above-described situation. To some extent, the effective
theory represents a general-relativistic generalization of v.
Seeliger's ansatz (\ref{3}).

\section{Conclusion} \label{ZUSAMMEN}

As introductorily mentioned, the idea of going beyond Riemannian
geometry can be motivated differently. In dependence on the chosen
approach, the question as to the matter coupled to gravitation has be
answered in a different manner. As long as one wants to reach a
unified geometric theory, there is no room left at all for any matter
sources. Since all matter was to be described geometrically such
sources are out of place. However, if one interprets the additional
geometric quantities as describing gravitation, then one has to
introduce matter sources. If one continues to assume the validity of
SEP in this case, one has to confine oneself to the consideration of
those purely metric theories which were reviewed in Sec.
\ref{4ORDNUNG}. In particular, this means that the source of
gravitation is assumed to be the metric energy-momentum tensor. If
one requires that the theory is derivable from a variation principle
this follows automatically. This mathematical automatism has a deep
physical meaning.

To make the latter point evident let us return to the discussion of
SEP. The formulation of SEP given above focuses its attention on the
coupling of gravitation to matter sources, without saying anything on
the structure of the matter sources. However, reminding the starting
point of the principle of equivalence, namely the Newtonian
equivalence between inertial, passive gravitational and active
gravitational masses, the SEP implies also a condition on the matter
source. Indeed, while the EEP is the relativistic generalization of
the equality of inertial and passive gravitational masses, the SEP
generalizes the equality of all three masses. Therefore, the
special-relativistic equivalent of the Newtonian inertial mass, i.e.,
the symmetrized special-relativistic energy-momentum tensor, has to
be lifted into the curved space and chosen as relativistic equivalent
of the active gravitational mass, that means, as source term in the
gravitational equations. Of course, this condition is automatically
satisfied if one starts from a Lagrangian in a Riemannian space.
Therefore, assuming that EEP  necessarily leads to a Riemannian
space, in \cite{54} this aspect of EEP was not mentioned explicitly.
But if one considers theories based on non-Riemannian geometry one
should keep in mind this condition required by SEP for the source
term.

As was shown in Sec. \ref{TETRAD}, the latter limitation on matter
can also be imposed on tetrad theories of gravity. However, it loses
its meaning if one goes over to more general geometries and theories
of gravity, respectively. In the case that the Lagrangian is given by
the Ricci scalar this even leads to a trivialization of the geometric
generalization \cite{138}. Therefore, it is more consequent to
consider non-Riemannian geometry and its perspectives for a
generalization of GRT from the standpoint of the matter sources, as
it was done in \cite{1191,1201,1211,1221,1231,HEHL}.

As the examples given above demonstrate, in post-Einsteinian theories
of gravity there are typical non-Einsteinian effects of gravitation
which correspond to certain pre-relativistic ansatzes. One meets
absorption (or/and self-absorption) and suppression (and/or
amplification) effects.

In case that the additions ($\Theta_{ik}$, $H_{ik}$, and $S_{Bki}$,
respectively) to the matter source, have a suitable sign there exists
an amplification of the matter source by its own gravitational field.
Above we called these terms "hidden- or dark-matter terms". To some
extent, this is justified by the following argument \cite{139}.

The fact that on large scales there is a discrepancy in the
mass-to-light ratios can be explained in two alternative ways. Either
one presupposes the validity of the theory of GRT and thus, in the
classical approximation, i.e., for weak fields (where $GM/rc^2\ll 1$)
and low velocities (where $v\ll c$), the validity of Newtonian
gravitational theory on large scales. Then one has to assume that
there exist halos of dark matter which are responsible for this
discrepancy. Or one takes a modification of GRT into consideration.
Then, the classical approximation of the modified theory should
represent a gravitational mechanics that is different from the
Newtonian one. For, since the dynamical determination of masses of
astrophysical systems is always performed in the classical
approximation, this approximate mechanics has to work with a
gravitational potential deviating from the usual $-GM/r$ form at
large distances. From this view larger mass parameters seem to be
responsible for the observed motions in astrophysical
systems\footnote{Another approach to a modified Newtonian law is
given by the MOND model. (For this see \cite{SAN02} and the
contribution of V. DeSabbata in this volume.)}.

Interestingly, the above-discussed pure gravitation-field additions
have one essential property of dark matter. They can be interpreted
as source terms whose divergence vanishes. Accordingly, they do not
couple to the optically visible matter and are themselves optically
invisible. Of course, to answer the question whether they can really
explain the astrophysical data one has to study the corresponding
solutions of the respective gravitational equations.

To summarize, when one presupposes the framework of GRT in order to
interpret the results of measurements or observations, then one finds
the following situation: In all three examples there are dark-matter
effects. In the case of tetrad theory, as an implication of the
potential-like coupling, additionally one finds a variable active
gravitational mass and a variable gravitational number $G$,
respectively. In the case of the Einstein-Cartan-Kibble-Sciama model
one finds a  $\Lambda$-term simulated by spinorial matter. Thus, if
one measures a variable $G$ number one has an argument in favor of a
theory with potential-like coupling. Moreover, dark-matter or
$\Lambda$-effects do not necessarily mean that one has to search for
exotic particles or that one has got an information about the value
of $\Lambda$ in GRT. These effects could also signal the need for a
transition to an alternative theory of gravity.

\begin{quote}

\end{quote}

\end{document}